\documentclass[aps,twocolumn,prd,superscriptaddress,preprintnumbers,showpacs]{revtex4}
\usepackage{graphicx}
\usepackage{bm}
\usepackage{amsmath}
\usepackage{amssymb}
\usepackage{amsthm}

\newcommand{\be}{\begin{equation}}
\newcommand{\ee}{\end{equation}}
\newcommand{\bea}{\begin{eqnarray}}
\newcommand{\eea}{\end{eqnarray}}
\newcommand{\bdm}{\begin{displaymath}}
\newcommand{\edm}{\end{displaymath}}
\newcommand{\beas}{\begin{eqnarray*}}
\newcommand{\eeas}{\end{eqnarray*}}
\def\dev#1#2{ {{\rm d} #1 \over {\rm d} #2} }

\begin{document}
\title{A stellar test of the physics of unification}

\author{J. P. P. Vieira}
\email[]{up090302017@alunos.fc.up.pt}
\affiliation{Centro de Astrof\'{\i}sica, Universidade do Porto, Rua das Estrelas, 4150-762 Porto, Portugal}
\affiliation{Departamento de F\'{\i}sica e Astronomia da Faculdade de Ci\^encias, Universidade do Porto, Rua do Campo Alegre, 4150-007 Porto, Portugal}
\author{C. J. A. P. Martins}
\email[]{Carlos.Martins@astro.up.pt}
\affiliation{Centro de Astrof\'{\i}sica, Universidade do Porto, Rua das Estrelas, 4150-762 Porto, Portugal}
\author{M. J. P. F. G. Monteiro}
\email[]{Mario.Monteiro@astro.up.pt}
\affiliation{Centro de Astrof\'{\i}sica, Universidade do Porto, Rua das Estrelas, 4150-762 Porto, Portugal}
\affiliation{Departamento de F\'{\i}sica e Astronomia da Faculdade de Ci\^encias, Universidade do Porto, Rua do Campo Alegre, 4150-007 Porto, Portugal}
\date{25 May 2012}
\begin{abstract}
We discuss the feasibility of using solar-type main-sequence stars as probes of fundamental physics and unification. We use a simple polytropic stellar structure model and study its sensitivity to variations of the gravitational, strong and electroweak coupling constants in the context of unification scenarios. We quantify the sensitivity of the Sun's interior temperature to these variations, finding $\left|\Delta\alpha/\alpha\right|\lesssim1.3\times10^{-4}$ for a 'canonical' choice of unification scenario, and discuss prospects for future improvements.
\end{abstract}

\maketitle

\section{\label{intro}Introduction}

Cosmology and particle physics are presently experiencing a truly exciting period. On the one hand, both have remarkably successful standard models \cite{Nakamura}, which are in agreement with a plethora of experimental and observational data. On the other hand, there are also strong hints that neither of these models is complete, and that new physics may be there, within the reach of the next generation of probes.

In this context, it is important to identify laboratory or astrophysical measurements that can give us more information about the nature and properties of this still unknown physics. Here we will discuss one such example. Several authors have suggested that stellar physics can be used to test the behavior of the gravitational sector \cite {Chang,Sakstein,Casanellas}. We will go further, as in \cite{neutron}, by showing how stellar physics can also provide information on the physics of unification and on the possible spacetime variation of nature's fundamental couplings.

Nature is characterized by a number of physical laws and fundamental dimensionless couplings, which historically we have assumed to be spacetime-invariant. For the former this assumption is a cornerstone of the scientific method, but for the latter it is an assumption with no justification \cite{DOV,RSoc}. There is ample experimental evidence showing that fundamental couplings run with energy, and many particle physics and cosmology models suggest that they should also roll with time \cite{Uzan}. This happens in cosmological models with dynamical scalar fields, including string theory.

Do the fundamental constants vary? Answering this question has key implications for cosmology and fundamental physics, and in particular can shed light on the enigma of dark energy \cite{Reconst0,Reconst2}. The current observational status is controversial, with claims for \cite{Webb,Reinhold,Dipole} and against \cite{Chand,King,Thompson} variations of the fine-structure constant $\alpha$ and the proton-to-electron mass ratio $\mu$ at redshifts $z\sim1-3$. As shown by recent workshops devoted to the topic the debate is open \cite{Book} and a resolution demands better data, but also independent ways to search for these variations, which may confirm or contradict these indications.

In addition to $\alpha$ and $\mu$, the gravitational constant $G$ is also interesting to study. Speaking of variations of dimensional constants has no physical significance: one can concoct any variation by defining appropriate units of length, time and energy. However, one can always choose an arbitrary dimensionfull unit as a standard and compare it with other quantities. If one assumes particle masses to be constant then constraints on $G$ are in fact constraining the (dimensionless) product of $G$ and the nucleon mass squared. A better route is to compare the strong interaction with the gravitational one: this can be done by assuming a fixed energy scale for Quantum Chromodynamics (QCD) and allowing a varying $G$, or vice-versa. With these caveats, probes of a rolling $G$ provide key information on the gravitational sector \cite{GBerro}.

Paradoxically, $G$ was one of the first constants to be measured (by Cavendish) but is now the least well known, a result of the weakness of gravity. In fact it is not known to better than percent level \cite{Will}. While a rolling $\alpha$ or $\mu$ implies a violation of the Einstein Equivalence Principle, a rolling $G$ is compatible with it, but it does violate the Strong Equivalence Principle. Any Grand-Unified Theory predicts a specific relation between variations of $\alpha$ and $\mu$ \cite{Coc}, and simultaneous measurements of both provide key consistency tests \cite{Reconst1}. Different classes of gravitational theories also predict different relations between these variations and those of $G$, so tighter gravitational constraints can discriminate between them.

Our work, which is in the same spirit as \cite{Adams,Ekstrom,neutron}, considers the feasibility of using stars as probes of fundamental physics and unification. Specifically, we are interested in the impact of possible spacetime variations of $\alpha$, $\mu$ and $G$ on the structure and evolution of stars. Stars are interesting for this purpose both because they are in principle sensitive to variations of all three relevant couplings and because their long lifetimes may make even tiny variations noticeable.

Presently we will limit ourselves to a study of polytropic stars. Although they are of limited use for describing realistic stars, they offer the advantage of mathematical simplicity, which in turn allows one to develop a clear physical picture of how the changes in the parameters impact the structure of the star. In subsequent work we will discuss more realistic stars, whose study requires detailed stellar evolution codes.

\section{\label{unify}Phenomenology of unification}

We wish to describe phenomenologically a broad class of models which allow for simultaneous variations of several fundamental couplings, such as the fine-structure constant $\alpha=e^2/(\hbar c)$, the proton-to-electron mass ratio $\mu=m_p/m_e$ and Newton's gravitational constant $G$. The simplest way to do this is to relate the various changes to those of a particular dimensionless coupling, typically $\alpha$. Then, if $\alpha=\alpha_0(1+\delta_\alpha)$ and
\begin{equation}
\frac{\Delta A}{A}=k_A\, \frac{\Delta\alpha}{\alpha} \,,
\end{equation}
we have $A=A_0(1+k_A\delta_A)$ and so forth.

The relations between the couplings are model-dependent. Here we will follow \cite{Coc}, considering a class of grand unification models in which the weak scale is determined by dimensional transmutation and further assuming that relative variation of all the Yukawa couplings is the same. Finally we assume that the variation of the couplings is driven by a dilaton-type scalar field \cite{Campbell}. In this case one find that the variations of $\mu$ and $\alpha$ are related through
\begin{equation}
\frac{\Delta\mu}{\mu}=[0.8~R-0.3~(1+S)]\frac{\Delta\alpha}{\alpha}\,,
\end{equation}
where $R$ and $S$ can be taken as free phenomenological parameters. Their absolute value can be anything from order unity to several hundreds, but while $R$ can be positive or negative (with the former case being more likely), physically one expects that $S>0$. Nevertheless, we can simply treat both as phenomenological parameters to be constrained by astrophysical data.

Concerning Newton's constant $G$, it is easiest to work in terms of the dimensionless couplings
\begin{equation}
\alpha_i=\frac{Gm^2_i}{\hbar c}\,.
\end{equation}
For our purposes it's natural to assume that the QCD scale and particle masses vary, while the Planck mass is fixed. We then have
\begin{equation}
\frac{\Delta\alpha_e}{\alpha_e}=2\frac{\Delta m_e}{m_e}
   =(1+S)~\frac{\Delta\alpha}{\alpha}
\end{equation}
and
\begin{equation}\label{alphap}
\frac{\Delta\alpha_p}{\alpha_p}=2\frac{\Delta m_p}{m_p}=2 \big[0.8~R+0.2 (1+S) \big]~\frac{\Delta\alpha}{\alpha}\,.
\end{equation}
Similarly for the mass difference between neutrons and protons, $\sigma=m_n-m_p$, we find
\begin{equation}
\frac{\Delta \sigma}{\sigma}=[0.1+0.7~S-0.6~R]\, \frac{\Delta\alpha}{\alpha}\,.
\end{equation}
For the ratio $\eta=m_n/m_p$, we have
\begin{equation}
\frac{\Delta \eta}{\eta}=\left(\frac{1}{\eta}-1\right)
   \left[0.1-0.5~S+1.4~R\right]\,\frac{\Delta\alpha}{\alpha}\,.
\end{equation}
Relative variations of quantities such as the neutron lifetime and the deuteron binding energy can also be cast in this form \cite{Coc}.

One important effect of varying masses is that they lead to a change in the star's density, even though the number density has the standard behavior. Indeed
\begin{equation}
\frac{\rho}{\rho_{\rm std}}=1+\frac{\Delta m_p}{m_p}=1+ \left[0.8~R+0.2 (1+S)\right]~\frac{\Delta\alpha}{\alpha}\,;
\end{equation}
thus the importance of this effect depends not only on the the value of the $\alpha$ variation but also on the values of the phenomenological parameters $R$ and $S$.

If one takes at face value the current evidence for a varying $\alpha$ \cite{Webb,Dipole} and the null results for $\mu$ \cite{King,Thompson}, one could infer that
\begin{equation}
\frac{\Delta\mu}{\mu} \ll \frac{\Delta\alpha}{\alpha}\,.
\end{equation}
If so, we can use the above $\mu$-$\alpha$ drift equation to obtain the following (approximate) relation
\begin{equation}
R\sim\frac{3}{8}\, (1+S)\,,
\end{equation}
which can in principle be used to simplify the analysis: in this case we have only two additional free parameters, $\Delta\alpha/\alpha$ and $R$. Although our analysis below will be generic, we will comment further on this particular limit. There are also \cite{Dipole} indications of a spatial $\alpha$ dipole, in the direction $RA=(17.3\pm1.0)h$, $\delta=(-61\pm10)^o$. If confirmed, this would suggest that stars in different directions would be subject to different couplings.

\section{\label{polytropic}Polytropic stars} 

A polytropic star (see \cite{Chandrasekhar}) is a simplified model for the structure of a star in equilibrium, built from the equation of mass and assuming hydrostatic equilibrium.
Specifically, it uses
\be
\dev{m}{r} = 4\pi r^2 \rho
\quad\quad {\rm and}\quad\quad
\dev{p}{r} = - {G m \rho \over r^2} \,.
\label{eq:eqs_structure}
\ee
Here $r$ is the radial distance to the centre of the star, $m$ the mass within the sphere of radius $r$, $\rho$ the density, and $p$ the pressure.
These equations can be solved if an equation of state of the type $p=p(\rho)$ is set.
The polytropic solution corresponds to a symplified equation of state in the form
\be
p = K \, \rho^{1 + 1/n} \,,
\ee
where $K$ is the polytropic constant (to be defined by the boundary conditions) and $n$ is the polytropic index (to be selected).
The use of this relation avoids the need to include an additional equation for temperature, describing the transport of energy in the interior.

In such a simplified approach the structure of the star is described by the numerical solution of the Lane-Emden Equation, as given by
\be
{ 1 \over \xi^2} \, \dev{ }{\xi}\left( \xi^2 \; \dev{\theta}{\xi} \right) = - \theta^n \;.
\label{eq:lane-emden}
\ee
In order to obtain this equation from Eqs~(\ref{eq:eqs_structure}), the following variables have been used; $\theta^n \equiv \rho/\rho_c$ and $\xi\equiv r/a$, with $\rho_c$ being the value of the density at the centre of the star and
\be
a^2 \equiv {K (n{+}1) \over 4\pi G \rho_c^{1-1/n}} \,.
\ee
The initial conditions, at the centre, for the Lane-Emden equation are $\theta(\xi=0)=1$ and $\theta^\prime(\xi=0)=0$.
The solution goes from $\xi=0$ to $\xi=\xi_s$, where $\theta(\xi_s) =0$ and $\theta^\prime_s\equiv\dev{\theta}{\xi}(\xi_s)<0$.
This value of $\xi_s$, corresponds to the first zero of the solution for $\theta$ for each value of $n$ used.
There is a finite value of $\xi_s\ge\sqrt{6}$ for any $0\le n < 5$.

If we consider a star with total mass $M_\star$ and radius $R_\star$ (not to be confused with the unification parameter $R$), then
\be
\rho_c = \left( - {\xi_s \over 3 \theta^\prime_s}\right) \, {3 M_\star \over 4\pi R_\star^3} \,,
\ee
and
\begin{equation}
K = {G \over n{+}1} \left[ {4\pi \over \xi_s^{n{+}1} (-\theta^\prime_s)^{n{-}1}} \; 
   {M_\star^{n{-}1} \over R_\star^{n{-}3}} \right]^{1/n}.
\end{equation}

The production of energy is not required, as any solution can be found with the selected radius and mass for a value of the index $n\in[0,5[$.
However, if the luminosity of the star ($L_\star$) is known, it may be used as a boundary condition that can establish the value of $n$.
Finally, if this boundary condition for the energy is used, then we can obtain the polytropic index $n$.
This is done by imposing that the luminosity of the polytrope is calculated from
\be
L_n \equiv {M \over \xi_s^2 \; (-\theta^\prime_s)} \int_0^{\xi_s} \xi^2 \theta^n \, \varepsilon(\xi) \, {\rm d}\xi,
\label{eq:luminosity}
\ee
where $\varepsilon$ is the emissivity, or energy production rate per unit mass and time, inside the star.
The expression for $\varepsilon$ will depend on the process used by the star to produce energy in the interior.
Therefore it will be given by the physics describing the fusion of hydrogen into helium in main sequence stars.
The version we adopt here are the global fits for the emissivity of the PP Chains (relevant for low mass stars, like the Sun or late-type stars).

When doing so, we need to assume an equation that provides the values of the temperature from density and pressure, for a known chemical composition. For the present work the ideal gas law is used, being an adequate approximation for stars, similar to the Sun, in the main sequence.
The value of $L_\star$ provides the condition to select $n$, if we require that the condition $L_n=L_\star$ is satisfied.

\section{Impact on the star from $\boldsymbol{\alpha}$} 

Since the variation of fundamental couplings directly affects the units we use to measure quantities with dimensions, it is convenient to work with dimensionless quantities. We will therefore use the dimensionless version of the previous equations by using the relations;
\begin{equation}
M_\star \rightarrow {M_\star\over m_p} ,\;\;
R_\star \rightarrow {R_\star \over m_{p}^{-1}\hbar^{1}c^{-1}},\;\;
\rho \rightarrow {\rho\over m_{p}^{4}\hbar^{-3}c^{3}}.
\end{equation}
Thus for density we have
\begin{equation}
{\rho_{c} \over m_{p}^{4}\hbar^{-3}c^{3}} =
   - {\displaystyle {M_\star / m_{p}} \over \displaystyle\left(R_\star \over m_{p}^{-1}\hbar^{1}c^{-1}\right)^3}\;
   {\xi_{s} \over 4\pi\theta_{s}^{\prime}} \,,
\label{eq:rho_c_a(n)}
\end{equation}
while for the polytropic constant
\begin{widetext}
\begin{equation}
{K \over m_{p}^{{-}4/n}\, \hbar^{3/n} \, c^{2-3/n}}=
    {\alpha_{p} \over n{+}1}\;
    \left[{\left(M_\star/m_{p}\right)^{n{-}1} \over \left(R_\star/m_{p}^{{-}1}\hbar^{1}c^{{-}1}\right)^{n{-}3}}\right]^{1/n}
    \left[\frac{4\pi}{\xi_{s}^{n+1}\left(-\theta_{s}^{\prime}\right)^{n-1}}\right]^{1/n} \,.
\label{eq:K_a(n)}
\end{equation}

For the emissivity, $\varepsilon_{ij}$, of the reaction between species $i$ and $j$, the following family of expressions can be used (see \citet{2011RvMP...83..195A} for details, and references therein);
\begin{equation}
{\varepsilon_{ij} \over m_{p}^{1}\hbar^{-1}c^{4}}=
   f \; \sqrt{8 \over 3A_{ij}} \;
   \left({\pi^{2}w_{ij} \over 2}\right)^{1/6} \;
   {S_{0} \over m_{p}^{-1}\hbar^{2}} \;
   {K^{-2/3}\rho_{c}^{1-2/3n} \over m_{p}^{4}\hbar^{-3}c^{5/3}} \;
   {\alpha^{1/3} \over \theta^{n+2/3}} \;
   \exp\left[-3 \;
   \left({\pi^{2}w_{ij} \over K \rho_{c}^{1/n} c^{-2}}\right)^{1/3}
   {\alpha^{2/3} \over \theta^{1/3}}\right]
\label{eq:emissividade (adimensional)}
\end{equation}
where $f$ is the energetic gain of the nuclear reaction in units of $m_pc^2$, $A_{ij}$ is the reduced atomic mass in units of $m_p$, $w_{ij}=z_iz_j A_{ij}$ where $z$ is the atomic number and $S_0$ the reference value for the S-factor (the effective cross section).
By using an expression for the emissivity, based on the dominant reaction of the PP Chain, we may then find the relation based on luminosity, that provides the relation that determines the value of $n$.
This is,
\begin{equation}
\frac{L_n}{m_{p}^{2}\hbar^{-1}c^{4}}=\left(4\pi\right)^{-\frac{3}{2}}\left(n+1\right)^{\frac{3}{2}}\left[\frac{\left(K\rho_{c}^{\frac{1}{n}-\frac{1}{3}}\right)^{\frac{3}{2}}}{m_{p}^{-2}\hbar^{\frac{3}{2}}c^{\frac{3}{2}}}\right]^{-\frac{3}{2}}\alpha_{p}^{-\frac{3}{2}}\intop_{0}^{\xi_{s}}\xi^{2}\theta^{n}\frac{\varepsilon}{m_{p}^{1}\hbar^{-1}c^{4}}d\xi\label{eq:luminosidade_a}\,.
\end{equation}
\end{widetext}

A model of a specific star, when we have $(M_\star,R_\star,L_\star)$, can be obtained by solving Eq.~(\ref{eq:lane-emden}) together with Eq.~(\ref{eq:luminosidade_a}), where we impose that $L_n=L_\star$.
The solution provides the behavior of $\rho$ and $p$.
Any change in the physics will imply a different solution in the interior for the same global parameters.
The model accommodates such a change by having a slightly different value of $n$ that fulfills the equation for the luminosity.

This approach allows us to obtain quantitative estimates of the relative change of the stellar structure when the underlying physics changes. Although a polytrope is not the best available model of a real star, it is a close enough model for our purposes.
The assumption behind this approach is that the relative change from the real star is of the same magnitude as the relative change of the polytropic model, when every other unknown aspect of the physics is not allowed to change.

\section{\label{modified}Results and constraints}

From observations of a specific star, the value of its stellar parameters may be obtained, specifically its mass $M_\star$, radius $R_\star$ and luminosity $L_\star$ (e.g. \citet{Creevey}, and references therein). This allows us to build a model, assuming that the set of equations and physics we use are valid, that fits these boundary conditions and describe the structure of the interior.
However if our physics is changed by a small amount, we must find a new model, adjusted to the same boundary conditions, but with a slightly different structure.
This approach can also be implemented for polytropic models, which allow us to quantify with great precision the change that one single aspect of the physics can have on the structure of the interior.
Specifically, we can determine by how much specific thermodynamic quantities will change in the interior when we introduce a small change in the physical constants entering the basic equations of stellar structure.

\begin{figure}
\includegraphics[width=9cm]{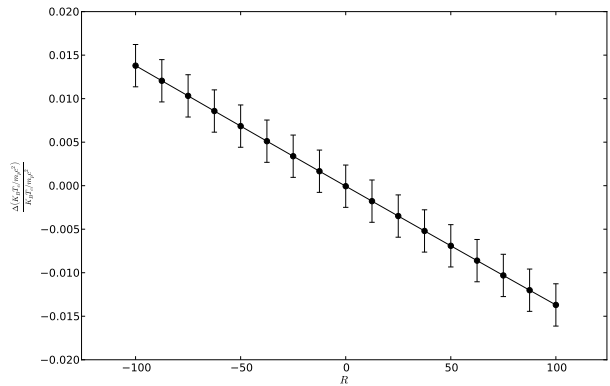}
\caption{\label{fig1}The relative change of the Sun's central temperature, for $\Delta\alpha/\alpha=-10^{-4}$, as a function of the unification parameter $R$ (with the choice $S=0$). Allowing a maximal variation of one percent yields the limit $\left|R\right|\lesssim90$.}
\end{figure}

\begin{figure}
\includegraphics[width=9cm]{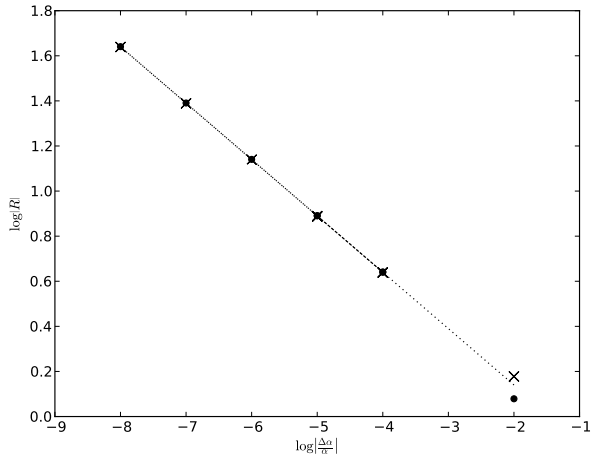}
\caption{\label{fig2}Bounds on the $(\alpha,R)$ parameter space, for the case $S=0$. The region above and to the right of the dashed line is excluded by our analysis. The dots (crosses) show the modulus of the estimated values of the maximum (minimum) allowed $R$ for particular choices of $\alpha$.}
\end{figure}

In order to have a useful test, we must refer to a quantity that can be obtained from the observation of real stars.
The ideal candidate is the value of the temperature near the centre, since this value has a direct effect on sound speed or neutrino production.
The former can be expected to be measured with significant precision with asteroseismology (or helioseismology for the solar case) while the latter may be used for the Sun.
We do not include the accumulated effect in time, since that requires a proper modeling of the evolution which is beyond the scope of the present work.

Presently, the precision we have for the temperature (or sound speed) in the interior of the Sun, is well below 1\% (see \cite{jcd}).
While far from this, a very high precision is also expected from the asteroseismology of other stars, as is currently being obtained with data from the NASA space mission Kepler (e.g. \citet{Chaplin}).

We thus use the formalism in the previous sections to estimate the impact that variations of $\alpha$ will have in the internal structure of a star for a given unification model (parametrized by specific values of $R$ and $S$).
We consider a polytropic star similar to our Sun (assumed to have a politropic index $n_0$), to which will correspond a slightly different value for $n$.
This star will have the same value of $M_\star/m_{p}$, $R_\star/(m_{p}^{-1}\hbar c^{-1})$, and $L_\star/(m_{p}^{2}\hbar^{-1}c^{4})$ ($M_\star$, $R_\star$ and $L_\star$ are the Sun's mass, radius and luminosity) but for a range of values where $\alpha$ differs from the standard value.
We then calculate by how much its central temperature (in fact the ratio $K_{B}T_{c}/(m_{p}c^2)$, where $T_{c}$ is the central temperature) must differ from that of our Sun, by solving Eq.~(\ref{eq:luminosidade_a}) to determine the value of $n$.

We have fixed specific values of $\Delta\alpha/\alpha$ and $S=0$ and let $R$ vary freely.
Since we are only considering the PP chains the main sensitivity of the polytropic model will be to changes in $\alpha_p$ (as well as $\alpha$ itself), and therefore we can only hope to constrain a linear combination of $R$ and $S$.
We may then assume $S=0$ and get a constraint for $R$, which from Eq. (\ref{alphap}) will in fact be a constraint on $R+S/4$. We have assumed $R_{min}\simeq-R_{max}$; in the range we have considered, $10^{-8}<\left|\Delta\alpha/\alpha\right|<10^{-2}$, this is an excellent approximation except at the top limit (where it is still adequate although no longer plausible).

In Fig. \ref{fig1} we show an example for $\Delta\alpha/\alpha=-10^{-4}$, in which case $R_{max}\simeq90$.
Repeating this analysis for different values of $\alpha$ we can identify the region in the $\left(R,S,\alpha\right)$ parameter-space which is consistent with the currently estimated experimental uncertainty for the central temperature of our Sun. This is summarized in Fig. \ref{fig2}, and yields our final bound
\begin{equation}\label{eq:bound}
\left|4R+S\right|\lesssim10^{-1.44}\left|\frac{\Delta\alpha}{\alpha}\right|^{-1}\,.
\end{equation}
As a simple illustration, if we take the typical values suggested in \cite{Coc} of $R\sim30$ and $S\sim160$, we find $\left|\Delta\alpha/\alpha\right|\lesssim1.3\times10^{-4}$.

\section{\label{concl}Conclusions}

We have used a simple polytropic model of stellar structure to quantify the impact of variations of the gravitational, strong and  electroweak coupling constants, in the context of unification scenarios, on solar-type stars. Although our analysis involves a number of simplifying assumptions, it serves to show that stars can yield tight constraints on these scenarios, with the added advantage that they nicely complement other astrophysical probes of these variations (such as quasar absorption systems and the cosmic microwave background) in terms of the characteristic density, length and time scales that they are sensitive to.

Naturally a polytropic model is of limited use for describing realistic stars.
Nevertheless, of particular relevance are stars near the lower limit for stellar mass, corresponding to fully convective stars (closely represented by a polytropic model with $n=1.5$) and having very long lifetimes in the main-sequence.
Asteroseismology of such stars may be able to provide robust limits on the central temperature, with this case corresponding very closely to the underlying assumptions used in this work.

The logical next step is to study more realistic models of stars, for which one requires detailed stellar evolution codes.
The very long lifetime of a star (particularly low-mass ones) is expected to be a key advantage since even tiny effects, accumulated over a long period, can have distinct observational consequences.

\begin{acknowledgments} 
This work was done in the context of projects PTDC/FIS/111725/2009 and PTDC/CTE-AST/098754/2008 from FCT, Portugal.
The work of JPV was partially funded by grant CAUP-02/2011-BII.  The work of CJM is
supported by a Ci\^encia2007 Research Contract, funded by FCT/MCTES (Portugal) and
POPH/FSE (EC), CJM and MJM are and also partially supported by grant PTDC/CTE-AST/098604/2008.
\end{acknowledgments}

\bibliography{stars}

\begin{thebibliography}{29}
\expandafter\ifx\csname natexlab\endcsname\relax\def\natexlab#1{#1}\fi
\expandafter\ifx\csname bibnamefont\endcsname\relax
  \def\bibnamefont#1{#1}\fi
\expandafter\ifx\csname bibfnamefont\endcsname\relax
  \def\bibfnamefont#1{#1}\fi
\expandafter\ifx\csname citenamefont\endcsname\relax
  \def\citenamefont#1{#1}\fi
\expandafter\ifx\csname url\endcsname\relax
  \def\url#1{\texttt{#1}}\fi
\expandafter\ifx\csname urlprefix\endcsname\relax\def\urlprefix{URL }\fi
\providecommand{\bibinfo}[2]{#2}
\providecommand{\eprint}[2][]{\url{#2}}

\bibitem[{\citenamefont{Nakamura et~al.}(2010)}]{Nakamura}
\bibinfo{author}{\bibfnamefont{K.}~\bibnamefont{Nakamura}} \bibnamefont{et~al.}
  (\bibinfo{collaboration}{Particle Data Group}), \bibinfo{journal}{J.Phys.G}
  \textbf{\bibinfo{volume}{G37}}, \bibinfo{pages}{075021}
  (\bibinfo{year}{2010}).

\bibitem[{\citenamefont{Chang and Hui}(2011)}]{Chang}
\bibinfo{author}{\bibfnamefont{P.}~\bibnamefont{Chang}} \bibnamefont{and}
  \bibinfo{author}{\bibfnamefont{L.}~\bibnamefont{Hui}},
  \bibinfo{journal}{Astrophys.J.} \textbf{\bibinfo{volume}{732}},
  \bibinfo{pages}{25} (\bibinfo{year}{2011}), \bibinfo{note}{* Temporary entry
  *}, \eprint{1011.4107}.

\bibitem[{\citenamefont{Davis et~al.}(2011)\citenamefont{Davis, Lim, Sakstein,
  and Shaw}}]{Sakstein}
\bibinfo{author}{\bibfnamefont{A.-C.} \bibnamefont{Davis}},
  \bibinfo{author}{\bibfnamefont{E.~A.} \bibnamefont{Lim}},
  \bibinfo{author}{\bibfnamefont{J.}~\bibnamefont{Sakstein}}, \bibnamefont{and}
  \bibinfo{author}{\bibfnamefont{D.}~\bibnamefont{Shaw}}
  (\bibinfo{year}{2011}), \bibinfo{note}{* Temporary entry *},
  \eprint{1102.5278}.

\bibitem[{\citenamefont{Casanellas et~al.}(2012)\citenamefont{Casanellas, Pani,
  Lopes, and Cardoso}}]{Casanellas}
\bibinfo{author}{\bibfnamefont{J.}~\bibnamefont{Casanellas}},
  \bibinfo{author}{\bibfnamefont{P.}~\bibnamefont{Pani}},
  \bibinfo{author}{\bibfnamefont{I.}~\bibnamefont{Lopes}}, \bibnamefont{and}
  \bibinfo{author}{\bibfnamefont{V.}~\bibnamefont{Cardoso}},
  \bibinfo{journal}{Astrophys.J.} \textbf{\bibinfo{volume}{745}},
  \bibinfo{pages}{15} (\bibinfo{year}{2012}), \eprint{1109.0249}.

\bibitem[{\citenamefont{P\`erez-Garc\'{\i}a and Martins}(2012)}]{neutron}
\bibinfo{author}{\bibfnamefont{M.~A.} \bibnamefont{P\`erez-Garc\'{\i}a}}
  \bibnamefont{and} \bibinfo{author}{\bibfnamefont{C.~J. A.~P.}
  \bibnamefont{Martins}} (\bibinfo{year}{2012}), \bibinfo{note}{* Brief entry
  *}, \eprint{1203.0399}.

\bibitem[{\citenamefont{Duff et~al.}(2002)\citenamefont{Duff, Okun, and
  Veneziano}}]{DOV}
\bibinfo{author}{\bibfnamefont{M.~J.} \bibnamefont{Duff}},
  \bibinfo{author}{\bibfnamefont{L.~B.} \bibnamefont{Okun}}, \bibnamefont{and}
  \bibinfo{author}{\bibfnamefont{G.}~\bibnamefont{Veneziano}},
  \bibinfo{journal}{JHEP} \textbf{\bibinfo{volume}{0203}}, \bibinfo{pages}{023}
  (\bibinfo{year}{2002}), \eprint{physics/0110060}.

\bibitem[{\citenamefont{Martins}(2002)}]{RSoc}
\bibinfo{author}{\bibfnamefont{C.~J. A.~P.} \bibnamefont{Martins}},
  \bibinfo{journal}{Phil.Trans.Roy.Soc.Lond.} \textbf{\bibinfo{volume}{A360}},
  \bibinfo{pages}{2681} (\bibinfo{year}{2002}), \eprint{astro-ph/0205504}.

\bibitem[{\citenamefont{Uzan}(2011)}]{Uzan}
\bibinfo{author}{\bibfnamefont{J.-P.} \bibnamefont{Uzan}},
  \bibinfo{journal}{Living Rev.Rel.} \textbf{\bibinfo{volume}{14}},
  \bibinfo{pages}{2} (\bibinfo{year}{2011}), \eprint{1009.5514}.

\bibitem[{\citenamefont{Nunes and Lidsey}(2004)}]{Reconst0}
\bibinfo{author}{\bibfnamefont{N.~J.} \bibnamefont{Nunes}} \bibnamefont{and}
  \bibinfo{author}{\bibfnamefont{J.~E.} \bibnamefont{Lidsey}},
  \bibinfo{journal}{Phys.Rev.} \textbf{\bibinfo{volume}{D69}},
  \bibinfo{pages}{123511} (\bibinfo{year}{2004}), \eprint{astro-ph/0310882}.

\bibitem[{\citenamefont{Amendola et~al.}(2011)\citenamefont{Amendola, Leite,
  Martins, Nunes, Pedrosa et~al.}}]{Reconst2}
\bibinfo{author}{\bibfnamefont{L.}~\bibnamefont{Amendola}},
  \bibinfo{author}{\bibfnamefont{A.~C.~O.} \bibnamefont{Leite}},
  \bibinfo{author}{\bibfnamefont{C.~J. A.~P.} \bibnamefont{Martins}},
  \bibinfo{author}{\bibfnamefont{N.~J.} \bibnamefont{Nunes}},
  \bibinfo{author}{\bibfnamefont{P.~O.~J.} \bibnamefont{Pedrosa}},
  \bibnamefont{et~al.} (\bibinfo{year}{2011}), \bibinfo{note}{* Temporary entry
  *}, \eprint{1109.6793}.

\bibitem[{\citenamefont{Murphy et~al.}(2004)}]{Webb}
\bibinfo{author}{\bibfnamefont{M.~T.} \bibnamefont{Murphy}}
  \bibnamefont{et~al.}, \bibinfo{journal}{Lect. Notes Phys.}
  \textbf{\bibinfo{volume}{648}}, \bibinfo{pages}{131} (\bibinfo{year}{2004}),
  \eprint{astro-ph/0310318}.

\bibitem[{\citenamefont{Reinhold et~al.}(2006)\citenamefont{Reinhold, Buning,
  Hollenstein, Ivanchik, Petitjean et~al.}}]{Reinhold}
\bibinfo{author}{\bibfnamefont{E.}~\bibnamefont{Reinhold}},
  \bibinfo{author}{\bibfnamefont{R.}~\bibnamefont{Buning}},
  \bibinfo{author}{\bibfnamefont{U.}~\bibnamefont{Hollenstein}},
  \bibinfo{author}{\bibfnamefont{A.}~\bibnamefont{Ivanchik}},
  \bibinfo{author}{\bibfnamefont{P.}~\bibnamefont{Petitjean}},
  \bibnamefont{et~al.}, \bibinfo{journal}{Phys.Rev.Lett.}
  \textbf{\bibinfo{volume}{96}}, \bibinfo{pages}{151101}
  (\bibinfo{year}{2006}).

\bibitem[{\citenamefont{Webb et~al.}(2011)\citenamefont{Webb, King, Murphy,
  Flambaum, Carswell et~al.}}]{Dipole}
\bibinfo{author}{\bibfnamefont{J.~K.} \bibnamefont{Webb}},
  \bibinfo{author}{\bibfnamefont{J.~A.} \bibnamefont{King}},
  \bibinfo{author}{\bibfnamefont{M.~T.} \bibnamefont{Murphy}},
  \bibinfo{author}{\bibfnamefont{V.~V.} \bibnamefont{Flambaum}},
  \bibinfo{author}{\bibfnamefont{R.~F.} \bibnamefont{Carswell}},
  \bibnamefont{et~al.}, \bibinfo{journal}{Phys.Rev.Lett.}
  \textbf{\bibinfo{volume}{107}}, \bibinfo{pages}{191101}
  (\bibinfo{year}{2011}), \eprint{1008.3907}.

\bibitem[{\citenamefont{Srianand et~al.}(2007)\citenamefont{Srianand, Chand,
  Petitjean, and Aracil}}]{Chand}
\bibinfo{author}{\bibfnamefont{R.}~\bibnamefont{Srianand}},
  \bibinfo{author}{\bibfnamefont{H.}~\bibnamefont{Chand}},
  \bibinfo{author}{\bibfnamefont{P.}~\bibnamefont{Petitjean}},
  \bibnamefont{and} \bibinfo{author}{\bibfnamefont{B.}~\bibnamefont{Aracil}},
  \bibinfo{journal}{Phys. Rev. Lett.} \textbf{\bibinfo{volume}{99}},
  \bibinfo{pages}{239002} (\bibinfo{year}{2007}).

\bibitem[{\citenamefont{King et~al.}(2008)\citenamefont{King, Webb, Murphy, and
  Carswell}}]{King}
\bibinfo{author}{\bibfnamefont{J.~A.} \bibnamefont{King}},
  \bibinfo{author}{\bibfnamefont{J.~K.} \bibnamefont{Webb}},
  \bibinfo{author}{\bibfnamefont{M.~T.} \bibnamefont{Murphy}},
  \bibnamefont{and} \bibinfo{author}{\bibfnamefont{R.~F.}
  \bibnamefont{Carswell}}, \bibinfo{journal}{Phys. Rev. Lett.}
  \textbf{\bibinfo{volume}{101}}, \bibinfo{pages}{251304}
  (\bibinfo{year}{2008}), \eprint{0807.4366}.

\bibitem[{\citenamefont{Thompson et~al.}(2009)}]{Thompson}
\bibinfo{author}{\bibfnamefont{R.~I.} \bibnamefont{Thompson}}
  \bibnamefont{et~al.}, \bibinfo{journal}{Astrophys. J.}
  \textbf{\bibinfo{volume}{703}}, \bibinfo{pages}{1648} (\bibinfo{year}{2009}),
  \eprint{0907.4392}.

\bibitem[{\citenamefont{Martins and Molaro}(2011)}]{Book}
\bibinfo{author}{\bibfnamefont{C.~J. A.~P.} \bibnamefont{Martins}}
  \bibnamefont{and} \bibinfo{author}{\bibfnamefont{P.}~\bibnamefont{Molaro}},
  \emph{\bibinfo{title}{{From varying couplings to fundamental physics}}}
  (\bibinfo{publisher}{Springer}, \bibinfo{year}{2011}).

\bibitem[{\citenamefont{Garcia-Berro et~al.}(2007)\citenamefont{Garcia-Berro,
  Isern, and Kubyshin}}]{GBerro}
\bibinfo{author}{\bibfnamefont{E.}~\bibnamefont{Garcia-Berro}},
  \bibinfo{author}{\bibfnamefont{J.}~\bibnamefont{Isern}}, \bibnamefont{and}
  \bibinfo{author}{\bibfnamefont{Y.~A.} \bibnamefont{Kubyshin}},
  \bibinfo{journal}{Astron.Astrophys.Rev.} \textbf{\bibinfo{volume}{14}},
  \bibinfo{pages}{113} (\bibinfo{year}{2007}).

\bibitem[{\citenamefont{Will}(2005)}]{Will}
\bibinfo{author}{\bibfnamefont{C.~M.} \bibnamefont{Will}},
  \bibinfo{journal}{Living Rev.Rel.} \textbf{\bibinfo{volume}{9}},
  \bibinfo{pages}{3} (\bibinfo{year}{2005}), \bibinfo{note}{an update of the
  Living Review article originally published in 2001}, \eprint{gr-qc/0510072}.

\bibitem[{\citenamefont{Coc et~al.}(2007)\citenamefont{Coc, Nunes, Olive, Uzan,
  and Vangioni}}]{Coc}
\bibinfo{author}{\bibfnamefont{A.}~\bibnamefont{Coc}},
  \bibinfo{author}{\bibfnamefont{N.~J.} \bibnamefont{Nunes}},
  \bibinfo{author}{\bibfnamefont{K.~A.} \bibnamefont{Olive}},
  \bibinfo{author}{\bibfnamefont{J.-P.} \bibnamefont{Uzan}}, \bibnamefont{and}
  \bibinfo{author}{\bibfnamefont{E.}~\bibnamefont{Vangioni}},
  \bibinfo{journal}{Phys.Rev.} \textbf{\bibinfo{volume}{D76}},
  \bibinfo{pages}{023511} (\bibinfo{year}{2007}), \eprint{astro-ph/0610733}.

\bibitem[{\citenamefont{Avelino et~al.}(2006)\citenamefont{Avelino, Martins,
  Nunes, and Olive}}]{Reconst1}
\bibinfo{author}{\bibfnamefont{P.~P.} \bibnamefont{Avelino}},
  \bibinfo{author}{\bibfnamefont{C.~J. A.~P.} \bibnamefont{Martins}},
  \bibinfo{author}{\bibfnamefont{N.~J.} \bibnamefont{Nunes}}, \bibnamefont{and}
  \bibinfo{author}{\bibfnamefont{K.~A.} \bibnamefont{Olive}},
  \bibinfo{journal}{Phys. Rev.} \textbf{\bibinfo{volume}{D74}},
  \bibinfo{pages}{083508} (\bibinfo{year}{2006}), \eprint{astro-ph/0605690}.

\bibitem[{\citenamefont{Adams}(2008)}]{Adams}
\bibinfo{author}{\bibfnamefont{F.~C.} \bibnamefont{Adams}},
  \bibinfo{journal}{JCAP} \textbf{\bibinfo{volume}{0808}}, \bibinfo{pages}{010}
  (\bibinfo{year}{2008}), \eprint{0807.3697}.

\bibitem[{\citenamefont{Ekstrom et~al.}(2010)\citenamefont{Ekstrom, Coc,
  Descouvemont, Meynet, Olive et~al.}}]{Ekstrom}
\bibinfo{author}{\bibfnamefont{S.}~\bibnamefont{Ekstrom}},
  \bibinfo{author}{\bibfnamefont{A.}~\bibnamefont{Coc}},
  \bibinfo{author}{\bibfnamefont{P.}~\bibnamefont{Descouvemont}},
  \bibinfo{author}{\bibfnamefont{G.}~\bibnamefont{Meynet}},
  \bibinfo{author}{\bibfnamefont{K.~A.} \bibnamefont{Olive}},
  \bibnamefont{et~al.}, \bibinfo{journal}{A. \& A.}
  \textbf{\bibinfo{volume}{514}}, \bibinfo{pages}{A62} (\bibinfo{year}{2010}),
  \eprint{0911.2420}.

\bibitem[{\citenamefont{Campbell and Olive}(1995)}]{Campbell}
\bibinfo{author}{\bibfnamefont{B.~A.} \bibnamefont{Campbell}} \bibnamefont{and}
  \bibinfo{author}{\bibfnamefont{K.~A.} \bibnamefont{Olive}},
  \bibinfo{journal}{Phys.Lett.} \textbf{\bibinfo{volume}{B345}},
  \bibinfo{pages}{429} (\bibinfo{year}{1995}), \eprint{hep-ph/9411272}.

\bibitem[{\citenamefont{{Chandrasekhar}}(1939)}]{Chandrasekhar}
\bibinfo{author}{\bibfnamefont{S.}~\bibnamefont{{Chandrasekhar}}},
  \emph{\bibinfo{title}{{An Introduction to the Study of Stellar Structure}}}
  (\bibinfo{publisher}{The University of Chicago Press}, \bibinfo{year}{1939}).

\bibitem[{\citenamefont{{Adelberger} et~al.}(2011)\citenamefont{{Adelberger},
  {Garc{\'{\i}}a}, {Robertson}, {Snover}, {Balantekin}, {Heeger},
  {Ramsey-Musolf}, {Bemmerer}, {Junghans}, {Bertulani}
  et~al.}}]{2011RvMP...83..195A}
\bibinfo{author}{\bibfnamefont{E.~G.} \bibnamefont{{Adelberger}}},
  \bibinfo{author}{\bibfnamefont{A.}~\bibnamefont{{Garc{\'{\i}}a}}},
  \bibinfo{author}{\bibfnamefont{R.~G.~H.} \bibnamefont{{Robertson}}},
  \bibinfo{author}{\bibfnamefont{K.~A.} \bibnamefont{{Snover}}},
  \bibinfo{author}{\bibfnamefont{A.~B.} \bibnamefont{{Balantekin}}},
  \bibinfo{author}{\bibfnamefont{K.}~\bibnamefont{{Heeger}}},
  \bibinfo{author}{\bibfnamefont{M.~J.} \bibnamefont{{Ramsey-Musolf}}},
  \bibinfo{author}{\bibfnamefont{D.}~\bibnamefont{{Bemmerer}}},
  \bibinfo{author}{\bibfnamefont{A.}~\bibnamefont{{Junghans}}},
  \bibinfo{author}{\bibfnamefont{C.~A.} \bibnamefont{{Bertulani}}},
  \bibnamefont{et~al.}, \bibinfo{journal}{Reviews of Modern Physics}
  \textbf{\bibinfo{volume}{83}}, \bibinfo{pages}{195} (\bibinfo{year}{2011}),
  \eprint{1004.2318}.

\bibitem[{\citenamefont{{Creevey} et~al.}(2007)\citenamefont{{Creevey},
  {Monteiro}, {Metcalfe}, {Brown}, {Jim{\'e}nez-Reyes}, and
  {Belmonte}}}]{Creevey}
\bibinfo{author}{\bibfnamefont{O.~L.} \bibnamefont{{Creevey}}},
  \bibinfo{author}{\bibfnamefont{M.~J.~P.~F.~G.} \bibnamefont{{Monteiro}}},
  \bibinfo{author}{\bibfnamefont{T.~S.} \bibnamefont{{Metcalfe}}},
  \bibinfo{author}{\bibfnamefont{T.~M.} \bibnamefont{{Brown}}},
  \bibinfo{author}{\bibfnamefont{S.~J.} \bibnamefont{{Jim{\'e}nez-Reyes}}},
  \bibnamefont{and} \bibinfo{author}{\bibfnamefont{J.~A.}
  \bibnamefont{{Belmonte}}}, \bibinfo{journal}{\apj}
  \textbf{\bibinfo{volume}{659}}, \bibinfo{pages}{616} (\bibinfo{year}{2007}).

\bibitem[{\citenamefont{{Christensen-Dalsgaard} and {Houdek}}(2010)}]{jcd}
\bibinfo{author}{\bibfnamefont{J.}~\bibnamefont{{Christensen-Dalsgaard}}}
  \bibnamefont{and} \bibinfo{author}{\bibfnamefont{G.}~\bibnamefont{{Houdek}}},
  \bibinfo{journal}{Astrop.Sp.Sc.} \textbf{\bibinfo{volume}{328}},
  \bibinfo{pages}{51} (\bibinfo{year}{2010}).

\bibitem[{\citenamefont{{Chaplin} et~al.}(2010)\citenamefont{{Chaplin},
  {Appourchaux}, {Elsworth}, {Garc{\'{\i}}a}, {Houdek}, {Karoff}, {Metcalfe},
  {Molenda-{\.Z}akowicz}, {Monteiro}, {Thompson} et~al.}}]{Chaplin}
\bibinfo{author}{\bibfnamefont{W.~J.} \bibnamefont{{Chaplin}}},
  \bibinfo{author}{\bibfnamefont{T.}~\bibnamefont{{Appourchaux}}},
  \bibinfo{author}{\bibfnamefont{Y.}~\bibnamefont{{Elsworth}}},
  \bibinfo{author}{\bibfnamefont{R.~A.} \bibnamefont{{Garc{\'{\i}}a}}},
  \bibinfo{author}{\bibfnamefont{G.}~\bibnamefont{{Houdek}}},
  \bibinfo{author}{\bibfnamefont{C.}~\bibnamefont{{Karoff}}},
  \bibinfo{author}{\bibfnamefont{T.~S.} \bibnamefont{{Metcalfe}}},
  \bibinfo{author}{\bibfnamefont{J.}~\bibnamefont{{Molenda-{\.Z}akowicz}}},
  \bibinfo{author}{\bibfnamefont{M.~J.~P.~F.~G.} \bibnamefont{{Monteiro}}},
  \bibinfo{author}{\bibfnamefont{M.~J.} \bibnamefont{{Thompson}}},
  \bibnamefont{et~al.}, \bibinfo{journal}{The Astroph. J. Let.}
  \textbf{\bibinfo{volume}{713}}, \bibinfo{pages}{L169} (\bibinfo{year}{2010}).

\end{thebibliography}

\end{document}